\documentclass[aps,preprint,showpacs,preprintnumbers,amsmath,amssymb]{revtex4}

\usepackage{graphicx}

\newcounter{saveeqn}
\newcommand{\alpheqn}{\setcounter{saveeqn}{\value{equation}}%
  \stepcounter{saveeqn}\setcounter{equation}{0}%
  \renewcommand{\theequation}{%
  \mbox{\arabic{saveeqn}\alph{equation}}}}%
\newcommand{\reseteqn}{\setcounter{equation}{\value{saveeqn}}%
\renewcommand{\theequation}{\arabic{equation}}}

\begin{document}

\title{Properties of fragmented repulsive condensates}
\author{Alexej I.\ Streltsov} 
\author{Lorenz S.\ Cederbaum} 
\affiliation{Theoretische Chemie,Universit\"at Heidelberg, D-69120 Heidelberg, Germany}
\date{\today}

\begin{abstract}
Repulsive Bose-Einstein condensates immersed into a double-well trap potential
are studied within the framework of the recently introduced mean-field approach which allows for bosons 
to reside in several different orthonormal orbitals.
In the case of a one-orbital mean-field theory (Gross-Pitaevskii) the ground state of the system
reveals a bifurcation scenario at some critical values of 
the interparticle interaction and/or the number of particles.
At about the same values of the parameters the two-orbital mean-field predicts 
that the system becomes two-fold fragmented.
By applying the three-orbital mean field 
we verify numerically that for the double-well external potential studied here the
overall best mean-field is achieved with two orbitals. 
The variational principle minimizes the energy at a vanishing population of the third orbital.
To discuss the energies needed to remove a boson from and 
the energies gained by adding a boson to the condensate, 
we introduce boson ionization potentials and boson affinities 
and relate them to the chemical potentials.
The impact of the finite number of bosons in the condensate on these quantities is analyzed.
We recall that within the framework of the multi-orbital mean-field theory
each fragment is characterized by its own chemical potential.
Finally, the stability of fragmented states is discussed in terms of the boson
transfer energy which is the energy needed to transfer a boson from one fragment
to another.
\end{abstract}

\maketitle

\section{ Introduction }
The experimental realization of Bose-Einstein condensation provides 
an unique opportunity to study macroscopic quantum systems.
One of the key property of a quantum system is the formation of shell structures.
For a system of $N$ identical fermions the formation of the shell structure is enforced by the Fermi-statistic, 
which implies a restriction to the number of particles residing in a single quantum level.
In contrast, for bosonic systems where the Bose-statistic permits any occupation numbers, 
the formation of shell structures, also known as the fragmentation phenomenon \cite{Nozieres},
must be caused by other reasons. 

Fragmentation can appear naturally in condensates made of different kinds of bosons.
Binary mixtures of trapped Bose-Einstein condensates have been the subject of numerous 
experimental \cite{Mix_exp1,Mix_exp2}
and theoretical investigations \cite{EsryGreene,PuBigelow,HoShenoy,CazalillaHo,SvidzinskyChui}.
These studies comprise binary mixtures made of two different alkalis such as 
$\,^{87}\!Rb-\,^{23}\!Na$ \cite{PuBigelow,HoShenoy}, 
of two different isotopes of the same atom  
$\,^{87}\!Rb-\,^{85}\!Rb$ \cite{EsryGreene,HoShenoy}, and
of two different hyperfine states of the same alkali such as the ($F=2,M_F=2$) and ($F=1,M_F=-1$)
states of $\,^{87}\!Rb$ \cite{Mix_exp1}. 

In contrast, the fragmentation phenomenon in BEC made of atoms
of one kind and in the same internal state, i.e., made of an identical bosons,
is an open theoretical frontier. 
Traditionally, BEC is described at the Gross-Pitaevskii (GP) mean-field level \cite{GP1,GP2}, 
where all bosons are residing in a single one-particle state. This one-orbital mean-field 
has been a very successful approximation and can explain many experiments;
see, e.g., Refs. \cite{Rev1,Rev2} and references therein.
However, this mean-field intrinsically cannot describe fragmentation.
The best mean-field approach allowing for bosons to reside in several different orthonormal orbitals
has been formulated recently \cite{BMF1,BMF2}. Each of the involved orbitals is characterized by a different spatial
distribution (localization in space), a different occupation number (number of bosons residing in the orbital),
and a different chemical potential.
This multi-orbital approach includes the Gross-Pitaevskii theory as a special case,
namely when all bosons reside in a single orbital.
The multi-orbital mean-field approach is based on a variational principle and hence
the optimal orbitals and their occupation numbers are determined variationally
such that the energy of the system is minimized.
In practice this means that the question whether a BEC forms a fragmented state (shell structure) or 
prefers to stay in a non-fragmented state (GP) is answered 
in the framework of one and the same method.

By applying this multi-orbital mean-field to study a repulsive BEC in multi-well external potentials
we found \cite{BMF3} that the ground state may be many-fold fragmented, 
i.e., the macroscopic occupation of several 
one-particle functions is energetically more favorable than the accumulation of all bosons in a single orbital. 
The quantities determining the fragmentation are the number of particles, the strength of 
the interparticle interaction and, of course, the specific shape of the external potential. 
The influence of all these parameters on fragmentation has been investigated in some detail.

In this paper we consider a system of $N$ identical bosons 
with positive scattering length immersed into a double-well external potential.
The Hamiltonian of this system is defined in Sec.II. 
In Sec.III we demonstrate that within the standard one-orbital mean-field theory (Gross-Pitaevskii),
the ground state of the system reveals a bifurcation scenario starting from a critical value of
the interparticle interaction strength or of the number of particles.
By applying two-orbital mean-field theory we show in Sec.IV that 
the ground state of the system made of $N$ identical bosons can indeed be two-fold fragmented.
The fragmentation phenomenon starts to take place in the vicinity of the 
bifurcation point obtained at the GP mean-field level.
In Sec.V we apply three-orbital mean-field theory 
and demonstrate that for the external potential studied here 
the overall best mean-field is achieved for two orbitals, i.e., 
the inclusion of a third orbital in the calculation does not improve the mean-field description
and the lowest energy is obtained if only two orbitals are occupied. 

Sec.VI opens a second part of the present investigation, where we 
introduce characteristic quantities which allow us to distinguish, at least in principle, between
fragmented and non-fragmented states of the condensates.
In particular, by comparing the systems of $N$ and $N\pm1$ bosons we adapt 
quantities relevant to fermionic systems
and define vertical and adiabatic ionization potentials and affinities for bosonic systems.
Their relevance to an experimental observation of fragmented states is also addressed.
In Sec.VII we introduce a boson transfer energy as the energy needed to transfer a boson from one
orbital to another one and use it to analyze the stability of fragmented states.
The discussion of the large N-limit of the boson ionization potential, 
boson affinity and boson transfer energy is presented in Sec.VIII.
Finally, Sec.IX summarizes our results and conclusions.

\section{The System and Hamiltonian}
Our general intention is to consider a system of $N$ identical bosons 
interacting via a $\delta$-function contact potential
$ W(\vec{r}_i-\vec{r}_j)=\lambda_0\,\delta(\vec{r}_i-\vec{r}_j) $, 
where $\vec{r}_i$ is the position of the i-th
boson and the nonlinear parameter $\lambda_0$ is related to the s-wave scattering 
length of the bosons \cite{Rev2}.
The Hamiltonian of this system takes on the standard form
\begin{equation}
\hat H =  
\sum_{i=1}^{N} \left[ -\frac{\hbar^2}{2m}\nabla^2_{\vec{r}_i} +
V(\vec{r}_i) \right] +  \sum_{i>j=1}^N W(\vec{r}_i-\vec{r}_j).
\label{hamiltonian}
\end{equation}
We denote as $h(\vec{r})=\hat{T}\,+\,V(\vec{r})$
the unperturbed one-particle Hamiltonian consisting of the
kinetic operator $ \hat{T} $ and the external potential $V(\vec{r})$. 

In this work we specifically study bosons trapped in the one-dimensional 
double-well external potential shown in the insets of Fig.\ref{fig1}.
Effectively the trap is obtained as an "inner" potential 
\begin{equation}
V_{inner}(x)= \omega (\frac{x^2}{2}-0.8)e^{-0.02(x^2+0.25x^3+6.1x)}
\label{InnerTrap}
\end{equation}
and an "outer" trap $V_{outer}$ which consists of an infinite wall at $x=9.5\pi$ where the inner potential
has already died off. The resulting combined potential ($V(x)\,=\,V_{inner}+V_{outer}$) 
has two well-separated nonequivalent wells.
It should be mentioned that the results discussed here hardly depend on whether the infinite wall of 
$V_{outer}$ is replaced by a smoothly growing potential wall \cite{BMF3}. 
The kinetic energy reads
$\hat{T}=-\frac{\omega}{2}\frac{\partial^2}{\partial x^2}$ implying that the coordinate $x$
is dimensionless while all energies and $\lambda_0$ are now in units of the frequency $\omega$.

We would like to stress that all conclusions and qualitative results discussed in this work
also apply to other double-well external potentials like, for instance, that 
used in the experimental set up of Ref.\cite{Markus}. 

\section{Gross-Pitaevskii Mean-Field Results} \label{GP}
The standard one-orbital mean-field description of the interacting system is obtained by assuming 
that the ground state
wave function $\Psi$ is a product of identical spatial orbitals $\varphi$:
$\Psi(\vec{r}_1,\vec{r}_2,\ldots,\vec{r}_N)= \varphi(\vec{r}_1)\varphi(\vec{r}_2)\cdots\varphi(\vec{r}_N)$.
The energy $E\,\equiv\,<\Psi|\hat{H}|\Psi>$, defined as the expectation value of the $\hat{H}$, reads
\begin{equation}
E_{GP}=N\{\int\varphi^* h\,\varphi\,d\vec{r}+\frac{\lambda}{2}\int|\varphi|^4\,d\vec{r}\},
\label{GP_energy}
\end{equation}
where $\lambda=\lambda_0(N-1)$ is the interaction parameter. 
By minimizing this energy the well-known Gross-Pitaevskii equation \cite{GP1,GP2} is obtained
\begin{equation}
\{\:h(\vec{r})+\lambda_0(N-1)|\varphi(\vec{r})|^2\}\,\varphi(\vec{r})=\,\mu_{GP}\,\varphi(\vec{r}).
\label{GP_orbital}
\end{equation}

The only parameter involved in the Gross-Pitaevskii mean-field is $\lambda=\lambda_0(N-1)$. 
Therefore, all systems which are characterized by the same $\lambda$
have the same energy per particle as well as the same orbital $\varphi(\vec{r})$ even if the systems 
have different numbers of bosons.
Below, we will use this fact to compare systems made of different numbers of bosons but have the same 
value of $\lambda$.

\begin{figure}
\includegraphics[width=11.2cm, angle=-90]{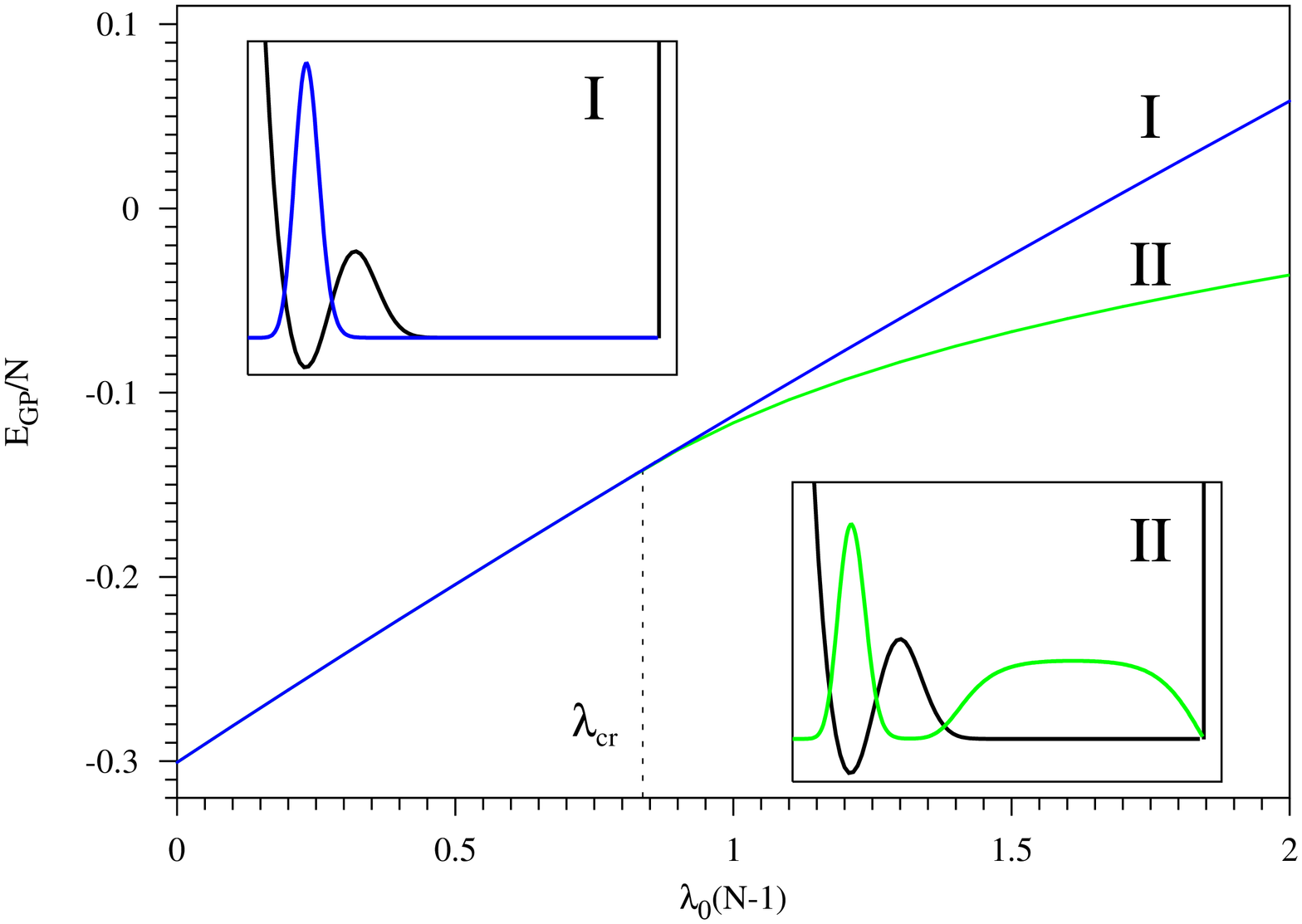}
\caption{(Color online) Results of one-orbital mean-field (GP) theory.
Shown are the energies per particle $E_{GP}/N$ of the lowest energy solutions of the Gross-Pitaevskii equation
as a function of $\lambda=\lambda_0(N-1)$. The bifurcation point observed at $\lambda_{cr}=0.837$
is marked by a dash line. In the insets the GP orbitals corresponding to the upper (I) 
and lower (II) energy branches are depicted together with the double-well trap potential. 
The values of the external trap potential (for parameters, see text) have been scaled by 1/10.
All energies are given in units of $\omega$. All orbitals are dimensionless and plotted as functions of
the dimensionless coordinate $x$.}
\label{fig1}
\end{figure}

We solved the Gross-Pitaevskii equation with the external potential introduced in Sec.II
for different values of $\lambda$.
In Fig.\ref{fig1} we plot the energy per particle of the lowest energy solutions 
as a function of $\lambda$. From this figure it can be inferred that 
the energy per particle of the ground state increases monotonically with $\lambda$ up to 
some critical value of $\lambda_{cr}=0.837$ and then the energy trajectory is split into two branches.
This critical value of $\lambda$ is indicated in Fig.\ref{fig1} by a vertical dashed line.  
The single-particle wavefunction (orbital) corresponding to the upper branch, marked as I in Fig.\ref{fig1},
is depicted in the upper inset of Fig.\ref{fig1}. 
This wavefunction is mainly localized in the left well.
From $\lambda_{cr}$ on, the lowest energy branch, labeled II in Fig.\ref{fig1}, smoothly 
bifurcates from the "localized" solution and describes
a delocalization of the BEC over the two wells. 
This delocalized solution is depicted in the right bottom inset 
of Fig.\ref{fig1}. For convenience, we also plot in both insets
the rescaled double-well potential $V/10$. 
The bifurcation scenario is a known 
feature \cite{BFRK1,BFRK2,BFRK3,BFRK4,BFRK5,CCI} of the Gross-Pitaevskii
equation reflecting the non-linearity of the underlying mean-field approximation. 
Bifurcations have been predicted for the ground \cite{BFRK1,BFRK2,BFRK4,BFRK5,CCI}
and excited \cite{BFRK2,BFRK3,BFRK5} states of repulsive \cite{BFRK2,BFRK3,BFRK4,BFRK5} 
and attractive \cite{BFRK1,BFRK2,BFRK4,BFRK5,CCI} condensates confined in different trap potentials.

Within the Gross-Pitaevskii mean field we obtain the following physical picture of Bose-Einstein 
condensation of repulsive bosons immersed into non-symmetric double-well potentials.
For $\lambda<\lambda_{cr}$ the BEC is localized in the deeper well only.
When $\lambda$ is increased, the BEC may continue to be localized in this well,
but it is energetically more favorable for bosons to tunnel through the barrier and 
to populate also the other well. 
As we have mentioned above, $\lambda=\lambda_0(N-1)$ is the only relevant parameter, and therefore both 
localization and delocalization phenomena can be observed for systems made of any number of particles
by properly tuning the interparticle interaction strength $\lambda_0$.

\section{Two-orbital Mean-Field Results}

In the previous section we demonstrated that above some critical value of nonlinearity $\lambda_{cr}$ 
the ground state of the system is described in the GP approach by a wave function delocalized over both wells.
The existence of a bifurcation in the GP results can be seen as a hint to go beyond the GP theory.
The natural question arises whether it is possible to improve the GP mean-field description
by providing a more flexible mean-field ansatz which would allow bosons to occupy 
two different orbitals which, in principle, could be localized in different wells.

Recently, a multi-orbital mean-field approach allowing for bosons to reside in
several different orthonormal one-particle functions has been formulated ~\cite{BMF1,BMF2}.
Using $m$ different orbitals, the approach is denoted MF($m$).
In this section we discuss MF(2).
In the next section we shall see that for the double-well geometry of the external potential studied here,
the overall best mean-field is achieved within the two-orbital mean-field theory.
Since the details of the derivations have already been published 
elsewhere ~\cite{BMF1,BMF2}, we present here only the final formulae.
Assuming that $n_1$ bosons occupy the orbital $\phi_1$ and $n_2$ bosons occupy the orbital $\phi_2$
leads to the following many-body wave function
\begin{eqnarray}
\Psi(\vec{r}_1,\ldots,\vec{r}_N)=
\hat{\cal S}\phi_1(\vec{r}_1)\cdots\phi_1(\vec{r}_{n_1})\phi_2(\vec{r}_{n_1+1})
\cdots\phi_2(\vec{r}_{n_1+n_2}),
\label{BMF2_ansatz}
\end{eqnarray}
where $\hat{\cal S} $ is the symmetrizing operator.
The MF(2) energy expression takes on the form
\begin{eqnarray}
 E = 
n_1 h_{11} + \lambda_0 \frac{n_1(n_1-1)}{2}\int|\phi_1|^4 d\vec{r}+ 
n_2 h_{22} + \lambda_0 \frac{n_2(n_2-1)}{2}\int|\phi_2|^4 d\vec{r}+ & & \nonumber\\
+2 \lambda_0 n_1 n_2 \int|\phi_1|^2 |\phi_2|^2 d\vec{r} 
\label{BMF2_energy}
\end{eqnarray}
The optimal orbitals $\phi_1$ and $\phi_2$ which minimize this energy functional are determined 
by solving the two coupled non-linear equations:
\begin{eqnarray}
\{\:h(\vec{r})+\lambda_0 (n_1-1)|\phi_1(\vec{r})|^2+2 \lambda_0 n_2|\phi_2(\vec{r})|^2  
\}\,
\phi_1(\vec{r}) =  \nonumber & & \\
= \mu_{11}\,\phi_1(\vec{r})+\mu_{12}\,\phi_2(\vec{r})  \nonumber & & \\
\{\:h(\vec{r})+\lambda_0 (n_2-1)|\phi_2(\vec{r})|^2+2 \lambda_0 n_1|\phi_1(\vec{r})|^2 
\}\,
\phi_2(\vec{r}) = \nonumber & & \\
= \mu_{21}\,\phi_1(\vec{r}) + \mu_{22}\,\phi_2(\vec{r})
\label{BMF2_orbital}
\end{eqnarray}

Obviously, the GP equations (\ref{GP_energy}) and (\ref{GP_orbital})
follow immediately from the MF(2) equations (\ref{BMF2_energy}) and (\ref{BMF2_orbital})
by putting the occupation of one of the two orbitals to zero.

In contrast to the one-orbital mean-field (GP) description where $\lambda=\lambda_0(N-1)$ is the only 
parameter involved, the two-orbital mean-field (MF(2)) depends on two parameters $n_1$ and $n_2$
which are the occupation numbers of the one-particle orbitals $\phi_1$ and $\phi_2$ respectively.
At fixed interaction strength $\lambda_0$ and number of bosons $N$,
the MF energy (see Eq.\ref{BMF2_energy}) depends on the particular value of $n_1$ ($n_2=N-n_1$).
It should be noted that the occupation numbers are variational parameters.
In order to find their optimal value and that of the energy, 
Eqs.(\ref{BMF2_orbital}) are solved for all possible occupation numbers, 
and the corresponding energies are evaluated by using Eq.(\ref{BMF2_energy}). 
At the optimal values of the occupation numbers the energy $E$ takes on its minimum and the MF(2)
becomes the best two-orbital mean field which we denote by BMF(2).

\begin{figure}
\includegraphics[width=11.2cm, angle=0]{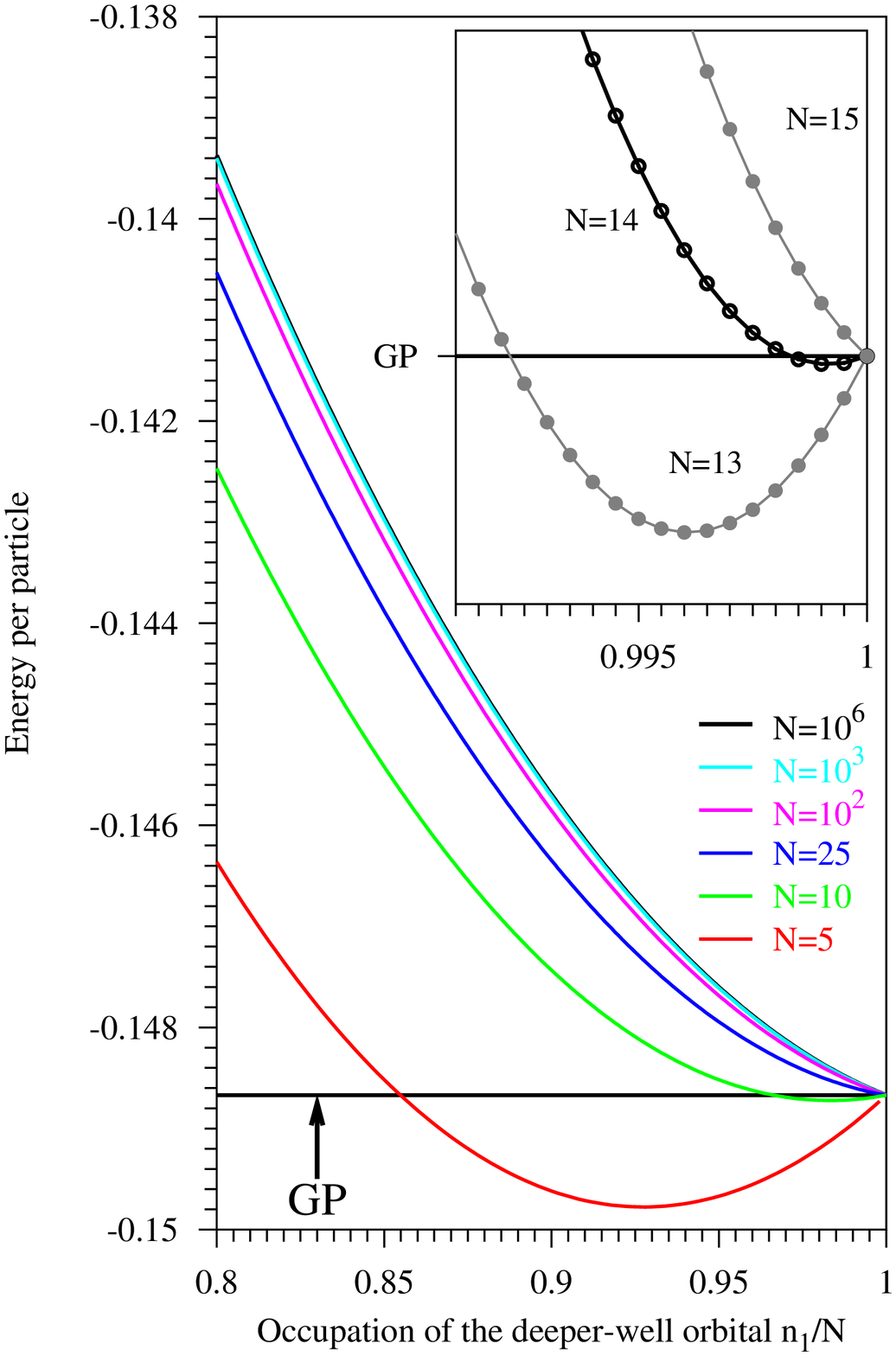}
\caption{(Color online) Results of the two-orbital mean-field (MF(2)) theory for $\lambda=0.8 <\lambda_{cr}$.
Shown are energies per particle for condensates made of $N=5,10,25,10^2,10^3$ and $10^6$ bosons
as a function of the relative occupation number $n_1/N$ of the orbital localized in the deeper well.
The horizontal solid line labeled as "GP" shows the corresponding GP energy per particle.
In the inset the energy per particle for the systems with $N=13,14$ and 15 bosons are plotted.
All energies are given in units of $\omega$.
}
\label{fig2A}
\end{figure}

\begin{figure}
\includegraphics[width=11.2cm, angle=0]{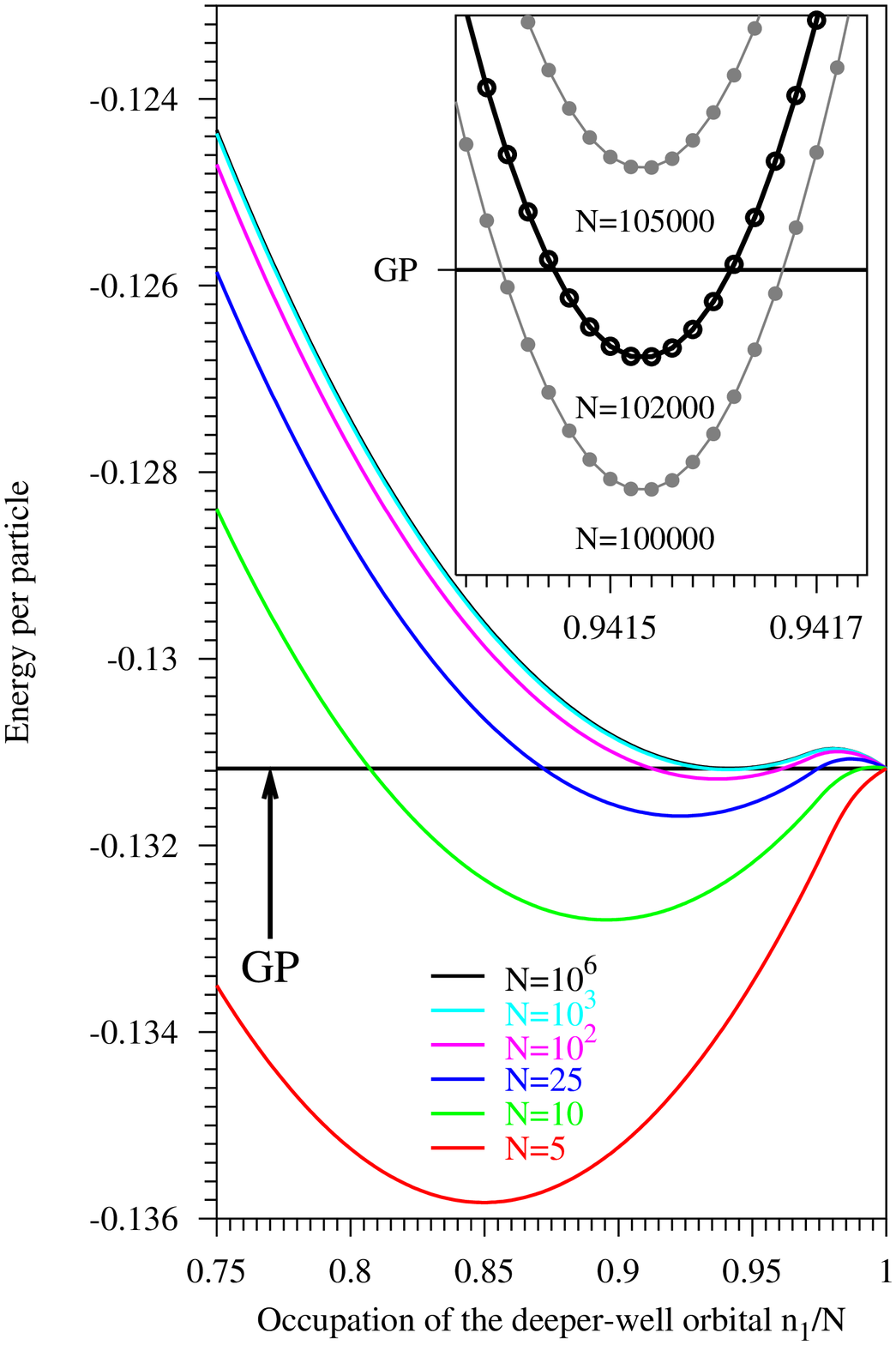}
\caption{(Color online) Results of the two-orbital mean-field (MF(2)) theory for $\lambda=0.9>\lambda_{cr}$.
Shown are energies per particle for condensates made of $N=5,10,25,10^2,10^3$ and $10^6$ bosons
as a function of the relative occupation number $n_1/N$ of the orbital localized in the deeper well.
The horizontal solid line labeled as "GP" shows the corresponding GP energy per particle.
In the inset the energy per particle for the systems with $N=100000,102000$ and 105000 bosons
are plotted.
All energies are given in units of $\omega$.
}
\label{fig2B}
\end{figure}

We solved the MF(2) equations for $\lambda=0.8$ and for $\lambda=0.9$.  
These values of the non-linear parameter have been chosen to be 
smaller and larger than $\lambda_{cr}=0.837$ which is the bifurcation
point of the GP energy trajectory discussed in Sec.III.
In Fig.\ref{fig2A} and Fig.\ref{fig2B} we plot
the MF(2) energies per particle for $\lambda=0.8$ and for $\lambda=0.9$,
respectively, 
as a function of the relative occupation number 
$n_1/N$ of the orbital localized in the deeper (left) well of the trap potential.
The results are shown for $N=5,10,25,10^2,10^3$ and $10^6$ bosons.
The GP energy is indicated by the horizontal line and labeled as "GP".

The common feature seen in Figs.\ref{fig2A} and \ref{fig2B} is that the energies per particle 
of the systems of bosons which are characterized by the same $\lambda=\lambda_0(N-1)$ 
do depend on the number of bosons $N$ and 
reveal a different dependence on $n_1/N$.
In contrast to the GP approach, the multi-orbital mean-field theory
distinguishes between systems with different number of bosons even if they are characterized by the same 
value of $\lambda$. As seen in the figures, for $n_1/N=1$, i.e., 
when all bosons reside in one orbital,
the energies of all the curves coincide at the
same value which is nothing but the corresponding GP energy. 
We can obviously conclude that the GP approach is indeed a special case of 
the MF(2) theory.

In Fig.\ref{fig2A} and in particular in Fig.\ref{fig2B} it is seen that the optimal BMF(2) energies may be 
lower than the corresponding GP one and in these cases BMF(2) constitutes the proper ground state of the system.
These ground states with nonzero occupation numbers are {\it fragmented} states,
because the respective one-particle density matrix has several nonzero eigenvalues \cite{Nozieres,Penrose,BMF3}.
Therefore, the fragmentation observed before in three-well potentials \cite{BMF3} also persists in double-well 
external potentials.
It is worthwhile to stress that at the minimum
of the energy one of the corresponding optimal orbitals is localized in the left well and the other one 
in the right well of the trap potential.
This is in an agreement with the three-fold fragmentation studied before, where the optimal
orbitals have also been found to be localized in the different wells 
of the three-well external potential \cite{BMF3}.

Figs.\ref{fig2A} and \ref{fig2B} exhibit several interesting features.
Let us first discuss Fig.\ref{fig2A} where $\lambda=0.8$ is smaller than $\lambda_{cr}$.
Inspection of the inset shows that the MF(2) energy curves as a function of $n_1/N$ posses
a minimum for $N\le14$ bosons. At these minima the energy is below the GP energy which is at
the maxima of the curves. For the systems which contain more than 14 particles, 
the MF(2) energy curves are monotonously increasing and their lowest energy is at the GP energy.

The situation changes drastically for $\lambda=0.9$ which is larger than $\lambda_{cr}$.
As seen in the inset of Fig.\ref{fig2B}, the BMF(2) energy is lower than the GP energy for
condensates with up to $N_{max}\approx 102000$ bosons. For all these systems the fragmentation
of the ground state is evident and macroscopic. One should be aware that $\lambda$ is kept
fixed in Fig.\ref{fig2B} and, therefore, the interparticle interaction strength $\lambda_0$ 
is small for the many-particle systems shown there.

Finally, we would like to point out two additional features of Fig.\ref{fig2B}.
First, the GP energy is a local minimum for most of the energy curves and consequently there 
appears a maximum which separates this energy from the global minimum, i.e., from the
BMF(2) energy. This may hint to an interesting dynamics of fragmentation if the system is
initially prepared in an non-fragmented state. 
Second, although there exists a maximal number of particles, $N_{max}$, for which
the ground state is fragmented, it should be stressed that for {\it all} $N>N_{max}$ 
the energy gap between the BMF(2) and GP energies is extremely small. 
For $N=10^6$ bosons, for instance, this energy difference per particle is only $\approx 3\times10^{-7}$
units of $\omega$. 
The coexistence of the fragmented and non-fragmented, but delocalized states is of
great interest by itself and might play a role in particular in time-dependent experiments.

\section{Proof that BMF(2) is the best overall mean field}

From the previous section we learned that 
by allowing bosons to reside in two orbitals,
the mean-field description of the repulsive BEC in the double-well potential 
has improved compared to the standard one-orbital description.
The natural question arises whether the inclusion of even more orbitals can provide 
further improvement of the mean-field description of the BEC.
To answer this question we apply the three-orbital mean-field theory (MF(3)) \cite{BMF3}.

The corresponding ansatz for the wavefunction assumes that 
three orbitals $\phi_1$, $\phi_2$ and $\phi_3$ are now occupied by
$n_1$,$n_2$ and $n_3=N-n_1-n_2$ bosons, respectively
\begin{eqnarray}
\Psi(\vec{r}_1,\ldots,\vec{r}_N)=
\hat{\cal S}\phi_1(\vec{r}_1)\cdots\phi_1(\vec{r}_{n_1})\phi_2(\vec{r}_{n_1+1})
\cdots\phi_2(\vec{r}_{n_1+n_2})\phi_3(\vec{r}_{n_1+n_2+1})
\cdots\phi_3(\vec{r}_{n_1+n_2+n_3}),
\label{BMF3_ansatz}
\end{eqnarray}
where $\hat{\cal S} $ is the symmetrizing operator.
The three-orbital mean-field energy reads \cite{BMF3}:
\begin{eqnarray}
 E =
n_1 h_{11} + \lambda_0 \frac{n_1(n_1-1)}{2}\int|\phi_1|^4 d\vec{r}+
n_2 h_{22} + \lambda_0 \frac{n_2(n_2-1)}{2}\int|\phi_2|^4 d\vec{r}+ & & \nonumber\\
n_3 h_{33} + \lambda_0 \frac{n_3(n_3-1)}{2}\int|\phi_3|^4 d\vec{r}+
+2 \lambda_0 n_1 n_2 \int|\phi_1|^2 |\phi_2|^2 d\vec{r} + & & \nonumber \\
+2 \lambda_0 n_1 n_3 \int|\phi_1|^2 |\phi_3|^2 d\vec{r} +
+2 \lambda_0 n_2 n_3 \int|\phi_2|^2 |\phi_3|^2 d\vec{r}.
\label{BMF3_energy}
\end{eqnarray}
The optimal orbitals minimizing this energy functional are obtained by solving the 
following system of three coupled non-eigenvalue equations \cite{BMF3}
\begin{eqnarray}
\{\:h(\vec{r})+\lambda_0 (n_1-1)|\phi_1(\vec{r})|^2+2 \lambda_0 n_2|\phi_2(\vec{r})|^2
+2 \lambda_0 n_3|\phi_3(\vec{r})|^2
\}\,
\phi_1(\vec{r}) =  \nonumber & & \\
= \mu_{11}\,\phi_1(\vec{r})+\mu_{12}\,\phi_2(\vec{r}) +\mu_{13}\,\phi_3(\vec{r}) \nonumber & & \\
\{\:h(\vec{r})+\lambda_0 (n_2-1)|\phi_2(\vec{r})|^2+2 \lambda_0 n_1|\phi_1(\vec{r})|^2
+2 \lambda_0 n_3|\phi_3(\vec{r})|^2
\}\,
\phi_2(\vec{r}) = \nonumber & & \\
= \mu_{21}\,\phi_1(\vec{r}) + \mu_{22}\,\phi_2(\vec{r})+\mu_{23}\,\phi_3(\vec{r}) \nonumber & & \\
\{\:h(\vec{r})+\lambda_0 (n_3-1)|\phi_3(\vec{r})|^2+2 \lambda_0 n_1|\phi_1(\vec{r})|^2
+2 \lambda_0 n_2|\phi_2(\vec{r})|^2
\}\,
\phi_3(\vec{r}) = \nonumber & & \\
= \mu_{31}\,\phi_1(\vec{r}) +\mu_{32}\,\phi_2(\vec{r}) + \mu_{33}\,\phi_3(\vec{r}).& &
\label{BMF3_orbital}
\end{eqnarray}
For the numerical procedure to solve this system of equations and to obtain 
the self-consistent orbitals $\phi_i$ and the corresponding values of the 
Lagrange parameters $\mu_{ij}$ ($i,j=1,2,3$), we refer to Ref.\cite{BMF3}.

\begin{figure}
\includegraphics[width=11.2cm, angle=-90]{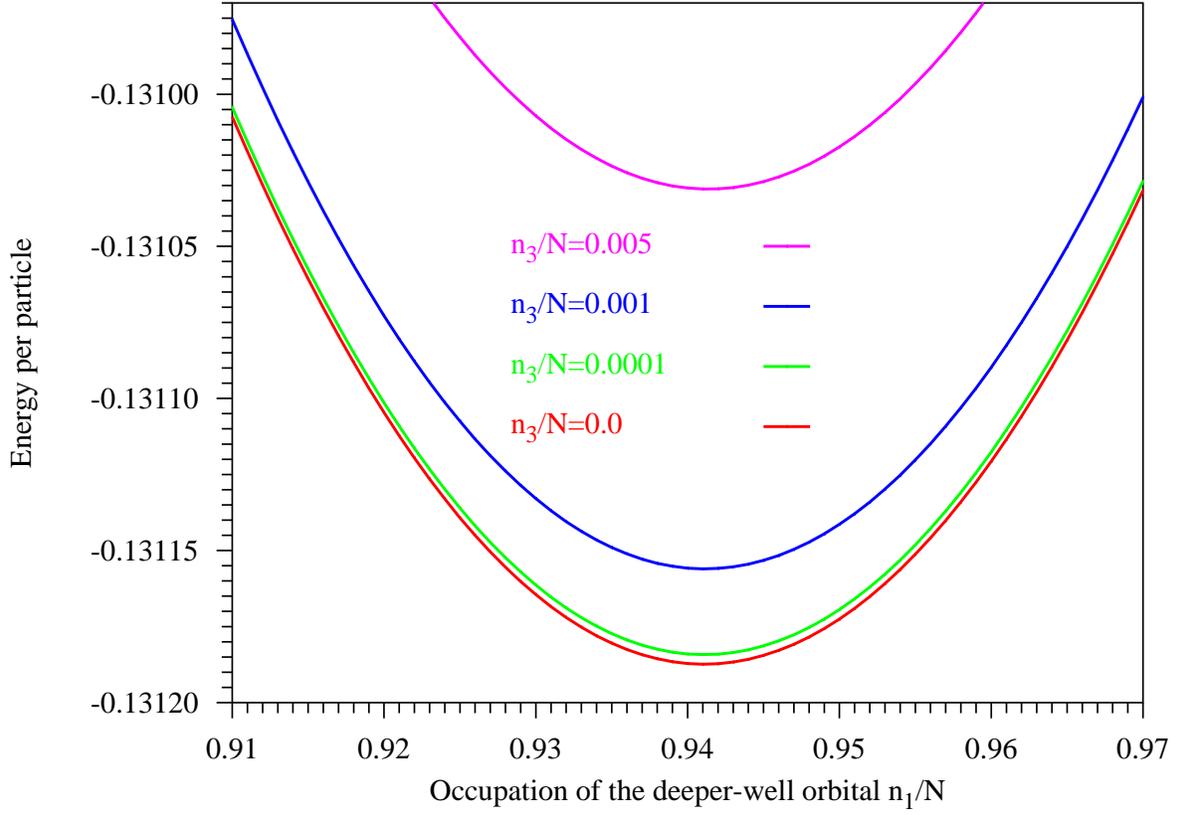}
\caption{(Color online)
Demonstration that BMF(2) provides the overall best mean field.
Shown are the results obtained with the 
three-orbital mean-field (MF(3)) theory for $N=1000$ bosons and $\lambda=\lambda_0(N-1)=0.9$.
Plotted are the energy per particle 
as a function of the relative occupation number $n_1/N$ 
for fixed values of the third relative occupation number $n_3/N$.
The curves shown correspond to $n_3/N = 0.0,0.0001,0.001$ and 0.005.
All energies are given in units of $\omega$.
}
\label{fig3}
\end{figure}

In comparison with the MF(2) approach the MF(3) method has one more variational parameter $n_3$.
We solved the above system of three coupled equations (\ref{BMF3_orbital}) 
for several values of $\lambda$ and $N$ and different fixed values of $n_3$.
As an example, we show in Fig.\ref{fig3} the results obtained 
for $\lambda=0.9$ and $N=1000$ for different occupation patterns.
Keeping the value of the third occupation number $n_3/N$ fixed at $0.0001,0.001,0.005$
and also at zero, we plot the three-orbital mean-field energy per particle as a function of
the relative occupation number $n_1/N$. 
The occupation of the second orbital $n_2=N-n_1-n_3$ is, of course, determined by $n_1$ and $n_3$.
The three-orbital mean-field method obviously includes MF(2) 
as a special case, namely, when the occupation of the third orbital vanishes. 
We see from Fig.\ref{fig3} that increasing the value of $n_3/N$ 
leads to a gradual increase of the MF(3) energy per particle.
Clearly, in the present case of repulsive condensates in the double-well potential, 
the overall best mean-field is obtained within the two-orbital BMF(2) theory. 
The enforced inclusion of more orbitals only enhances the energy of the condensate.

Several consequences should be mentioned.
First, fragmentation is a general physical phenomenon which takes place in repulsive BECs trapped 
in multi-well external potentials. 
Second, the number of the fragments and their occupation numbers are defined variationally, 
by minimizing the total energy functionals in Eqs.(\ref{BMF2_energy}) and (\ref{BMF3_energy}). 
If more orbitals are included in the mean-field ansatz than needed, 
the occupation of the superfluous orbitals becomes zero.

\section{Some properties of fragmented states}\label{BIP_BA}

Usually, to distinguish fragmented and non-fragmented states 
a phase difference between wave functions of the different fragments is considered \cite{phase1,phase2,phase3}.
Dynamical stability \cite{ref1,ref2} of the relative phase between the condensates localized 
in the different wells and related questions on the evolution of the fragmented state \cite{JJ,JJ1} 
have been a subject of several discussions.

In contrast to these time-dependent studies on fragmentation, 
we concentrate in the second part of our work on time-independent properties 
of the stationary fragmented states themselves.
The analogy between shell structures (atoms and molecules) formed in the fermionic world and the 
bosonic fragmented states studied in the previous sections motivates us to make use of 
some fermionic observables such as the ionization potential and electron affinity, 
and introduce related quantities for bosonic systems. 
Indeed, at the multi-orbital mean-field level of description the fermionic system is described by 
antisimmetrized single-determinant wave function \cite{CCTB}, while the 
symmetrized ansatz (see Eqs.(\ref{BMF2_ansatz}) and (\ref{BMF3_ansatz})) is proposed for 
bosonic systems. The variationally optimal orbitals and the corresponding orbital energies are obtained by solving 
the well-known Hartree-Fock (HF) equations \cite{CCTB} for fermions and the BMF equations for bosons.
The total energies of these systems are evaluated by computing the expectation values 
of the full many-body Hamiltonians with the (anti)symmetrized ansatz of the wave function.

The electron ionization potential (IP) is defined as the difference in total energies between
the reference system with $N$ electrons and the ionized system with $N-1$ electrons \cite{CCTB}. 
Because of this definition, the ionization potential is also referred to 
as the binding energy of an electron in the system.
We define the {\it boson ionization potential} or the binding energy of a boson 
as the energy needed to remove a boson from a bosonic system with $N$ bosons.  
Another very important physical characteristic is the electron affinity defined 
as the energy gained by adding one electron to the atom or molecule \cite{CCTB}.
In analogy, we define the {\it boson affinity } as the energy gained 
by attaching one boson to the bosonic system under consideration.

The boson ionization potentials and affinities can be calculated straightforwardly 
based on their definitions as differences between total energies.
In principle, these total energies may be evaluated within the framework of any suitable N-body 
method, but in the present study we restrict ourselves for consistency to the two-orbital mean-field approach.
To distinguish between the total energies of the condensates made of a different number of bosons 
we introduce superscript indices. In this notation the MF(2) total energies 
of the system with $N-1,N$ and $N+1$ bosons take on the following form
$^{N-1}\!E(n_1,n_2),\,^{N}\!E(n_1,n_2),\,^{N+1}\!E(n_1,n_2)$, where the superscript   
refers to the condensates made of the corresponding number of bosons.
We recall that within the framework of the MF(2) approach the total energy of the system 
is a function of the occupation number $n_1$ ($n_2=N-n_1$) 
and the minimum of this energy is called the best mean field (BMF(2)).
Obviously, the MF(2) energies of the systems with a different number of bosons 
achieve their minimal values (BMF(2) energies) at different values of the occupation numbers.  
We denote these BMF(2) energies and corresponding occupation numbers as $^{K}\!E_0$ and $(n^K_1,n^K_2)$ 
where the superscript $K$ refers to the system with $K$ bosons.


\subsection{Adiabatic and Vertical boson ionization potentials}

The general definition of the boson ionization potential 
involves the total energies of condensates made of $N$ and $N-1$ bosons.
Usually, the reference system with $N$ bosons is considered 
to be in its ground equilibrium state while the ionized system with $N-1$ bosons
can be either in an equilibrium or in a some transitional state.
In the following section we verify that only the BMF(2) energy (the minimum of the MF(2) energy curve)
corresponds to the equilibrium state of the condensate, while all other MF(2) energies 
can be attributed to some transitional (non-equilibrium) states of the condensate.

The difference between the ground {\it equilibrium} state energies of the ionized system with $N-1$ bosons
and the reference system with $N$ bosons is called {\it adiabatic boson ionization potential} ($BIP_A$):
\begin{equation}
BIP_A\,=\,^{N-1}\!E_0-\,^{N}\!E_0.
\label{BIP_A}
\end{equation}
The equilibrium in the ionized state might be achieved by adiabatically removing a boson from the condensate.

In contrast, if a boson is suddenly removed from the system, then
the state created is not an equilibrium state of the ionized system and
the difference between the energy of this non-equilibrium state  
and that of the ground state of the reference system is called 
{\it vertical boson ionization potential} ($BIP_V$).
If the bosonic system under consideration is fragmented, then
the sudden ionization of a boson from different orbitals of the fragmented state
requires different energies. The two-fold fragmented states are characterized by two 
different vertical boson ionization potentials
\begin{eqnarray}
BIP_V(1)\,=\,^{N-1}\!E(n^N_1-1,n^N_2)-\,^{N}\!E_0(n^N_1,n^N_2)\,\nonumber\\
BIP_V(2)\,=\,^{N-1}\!E(n^N_1,n^N_2-1)-\,^{N}\!E_0(n^N_1,n^N_2),\nonumber\\
\label{BIP_V}
\end{eqnarray}
where $^{N}\!E_0\,=\,^{N}\!E_0(n^N_1,n^N_2)$ is the BMF(2) energy of the reference system with $N$ bosons 
obtained at the optimal occupation numbers $n^N_1$ and $n^N_2$ of the two fragments and,
$^{N-1}\!E(n^N_1-1,n^N_2)$ and $^{N-1}\!E(n^N_1,n^N_2-1)$ are the MF(2) energies of the ionized states 
where a boson has been removed from the first and second fragment, respectively.

\begin{figure}
\includegraphics[width=11.2cm, angle=-90]{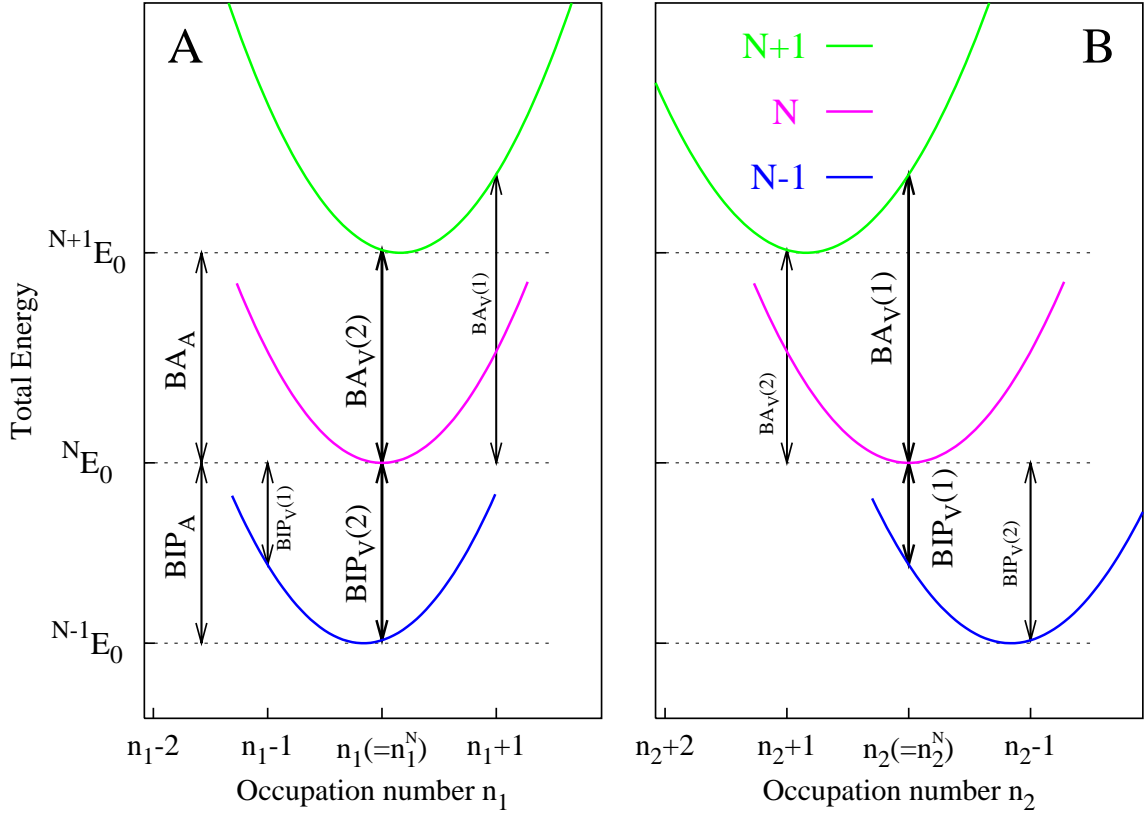}
\caption{(Color online) Schematic diagram defining the boson ionization potentials (BIP) and boson affinities (BA).
Total energies of condensates with $N-1,N$ and $N+1$ bosons obtained within the framework of the MF(2) theory
are plotted in the left figure (A) as a function of the occupation number $n_1$ and
in the right figure (B) as a function of the occupation number $n_2=N-n_1$.
The minima of the energy curves are denoted by $\,^{N-1}\!E_0,\,^N\!E_0$ and $\,^{N+1}\!E_0$,
and are indicated on the energy axis.
The reference condensate for which the boson ionization potentials and boson affinities are defined is
that with $N$ particles. The optimal occupation numbers of this condensate are $n^N_1$ and $n^N_2$.
Vertical and adiabatic boson ionization potentials and boson affinities are shown (see text for more details).
}
\label{fig4}
\end{figure}

In Fig.\ref{fig4}A we present a schematic diagram 
of the adiabatic and vertical ionization potentials together with 
the MF(2) total energies of the systems with $N+1,N$ and $N-1$ bosons 
plotted as functions of the number $n_1$ of bosons residing in the first orbital.
In Fig.\ref{fig4}B we plot these quantities as functions of the complementary 
parameter $n_2=N-n_1$ (the number of bosons residing in the second orbital).
The BMF(2) energies of the systems with $N+1,N$ and $N-1$ bosons are indicated on the energy axis.
The optimal occupation numbers $n^N_1$ and $n^N_2$ corresponding to the minimum of the MF(2) 
energy of the reference N-boson system are indicated on the x-axes of Fig.\ref{fig4}A and Fig.\ref{fig4}B,
respectively.

The difference between the ground state energy of the ionized system $^{N-1}\!E_0$
and that of the reference system $^{N}\!E_0$, i.e.,
the adiabatic ionization potential, is marked in Fig.\ref{fig4}A as $BIP_A$.
The vertical boson ionization potential $BIP_V(2)$ 
is shown in Fig.\ref{fig4}A as a {\it vertical} line
connecting the minimum of the $N$-particle energy curve 
and the energy curve of the system with $N-1$ bosons.
Let us explain this construction. When a boson is suddenly removed from the 
second orbital $\phi_2$ of the reference system, 
the occupation number $n^N_1$ of the other (first) orbital remains the same also 
in the ionized system, while 
the occupation of the second orbital in the ionized state is obviously reduced by one to $n^N_2-1$.
Therefore, this ionization process is indicated by the vertical line 
in the $(E,n_1)$-diagram.

The vertical boson ionization potential $BIP_V(1)$ is also shown as 
a vertical line connecting the minimum of the N-particle energy curve 
and the energy curve of the system with $N-1$ bosons, but in the complementary
$(E,n_2=N-n_1)$-diagram (see Fig.\ref{fig4}B). In this case the occupation number of the
first orbital is reduced by one to $n^N_1-1$ in the ionized state, while 
the occupation number of the second orbital $n^N_2$ remains the same as in 
the reference state.
For completeness, the $BIP_V(1)$ and $BIP_V(2)$ are plotted in both panels of figure \ref{fig4}. 

Removing a single boson from the system will not cause a strong 
change of the orbitals of the other bosons and these orbitals can be assumed fixed. 
In fermionic systems this assumption is known as the {\it frozen orbital approximation} \cite{CCTB}. 
This observation allows us to evaluate {\it approximately} the vertical ionization potentials 
and hence, the total energies of the ionized systems.

In the frozen orbital approximation we assume that the orbitals of the reference N-bosonic system 
$\phi_1$ and $\phi_2$ do not change upon the sudden removal of a boson, i.e.,
the ionized system is described by the same orbitals as well.
The energies of the corresponding ionized states where the boson has been suddenly removed from 
the $\phi_1$ or the $\phi_2$ orbital take on the form 

\alpheqn\begin{eqnarray}
^{N-1}\!E(n_1-1,n_2) \,=\, (n_1-1) h_{11}  
+\lambda_0 \frac{(n_1-1)(n_1-2)}{2}\int|\phi_1|^4 d\vec{r} + & & \nonumber\\
n_2 h_{22}  +  \lambda_0 \frac{n_2(n_2-1)}{2}\int|\phi_2|^4 d\vec{r} 
 +  2 \lambda_0 (n_1-1) n_2 \int|\phi_1|^2 |\phi_2|^2 d\vec{r} \label{E_N-1.A} \\ \nonumber\\
^{N-1}\!E(n_1,n_2-1) \,=\, (n_2-1) h_{22} 
+ \lambda_0 \frac{(n_2-1)(n_2-2)}{2}\int|\phi_2|^4 d\vec{r} + & & \nonumber\\
n_1 h_{11}  +  \lambda_0 \frac{n_1(n_1-1)}{2}\int|\phi_1|^4 d\vec{r}+ 
 + 2 \lambda_0 n_1 (n_2-1) \int|\phi_1|^2 |\phi_2|^2 d\vec{r}  \label{E_N-1.B}
\end{eqnarray}
\reseteqn
respectively.

According to the definitions in Eqs.\ref{BIP_V},
the vertical ionization potentials are obtained 
by substructing the BMF energy $^{N}\!E_0$ of the reference N-particle state 
from the energies of the ionized states in Eqs.\ref{E_N-1.A},\ref{E_N-1.B} at $n_2=n^N_2$.
We easily find for the boson vertical ionization potentials in the frozen orbital 
approximation the appealing relations 
\begin{eqnarray}
BIP_V(1)&=&-\mu_{11}\nonumber\\
BIP_V(2)&=&-\mu_{22}, 
 \label{BIP_V_FRZ}
\end{eqnarray}
i.e., they are given by the negative of the respective Lagrange multipliers which 
we call chemical potentials.

It is worthwhile to recall that at the level of the standard GP theory (see Sec.\ref{GP})
the energy needed to remove a boson from the non-fragmented N-boson condensate 
without changing the corresponding orbital $\varphi$ is given by the GP chemical potential: 
\begin{equation}
\,^{N-1}\!E_{GP}-\,^N\!E_{GP}\,=\,-\mu_{GP}.
\label{IP_GP}
\end{equation}

The relations between orbital energies and ionization potentials in the fermionic case 
are the subject of the well-known Koopmans' theorem \cite{CCTB,Koopmans}. Whithin the Hartree-Fock
mean field this theorem states that the vertical ionization potential is 
the negative of the energy of the orbital from which the electron has been removed. 
Attributing the diagonal Lagrange multipliers (chemical potentials) $\mu_{ii}$ to 
the orbital energies of the fragmented state, the formal results for 
fermionic and bosonic systems are absolutely identical.

We conclude that the process of sudden ionization 
of a boson from the non-fragmented GP state is characterized by a single
ionization potential while different ionization potentials
characterize a two-fold fragmented state of the condensate.
We recall that in the thermodynamic limit the change in the energy of a system 
obtained by removing a particle is called {\it chemical potential}.
The condensates studied here are made of identical bosons, 
and, therefore, a single chemical potential is expected to characterize these condensates
at equilibrium.
This seems to contradict the results obtained here for the two-fold fragmented states, where
two, in general different, chemical potentials exist.
This aparent contradiction is resolved by remembering that 
the present results are for finite systems. Indeed, we demonstrate 
in Sec.\ref{LN} that in the limit of a large number of particles
these different chemical potentials become identical at the optimal occupations
restoring thereby the thermodynamic picture of a condensate.

\subsection{Adiabatic and vertical boson affinities.}
The main propose of the present section is to consider 
a process where a boson is added to the reference system of $N$ bosons.
We define {\it adiabatic boson affinity} ($BA_A$) 
as the difference between the ground {\it equilibrium} state energies
of this reference system and of the system which results by attaching a boson to the reference system
\begin{equation}
BA_A\,=\,^{N}\!E_0-\,^{N+1}\!E_0.
\label{BA_A}
\end{equation}
The adiabatic boson affinity is shown schematically in Fig.\ref{fig4}A. 
It describes the energy gained by the attachment of a boson.

Similarly to the sudden ionization process used to introduce the vertical 
ionization potentials in the previous subsection,
we consider the sudden attachment of a boson to the two-fold fragmented state 
and define two {\it vertical boson affinities} comprising the 
sudden attachment of a boson to the $\phi_1$ and $\phi_2$ fragments, respectively:
\begin{eqnarray}
BA_V(1)\,=\,^{N}\!E_0(n^N_1,n^N_2)-\,^{N+1}\!E(n^N_1+1,n^N_2)\nonumber\\
BA_V(2)\,=\,^{N}\!E_0(n^N_1,n^N_2)-\,^{N+1}\!E(n^N_1,n^N_2+1). \nonumber\\
\label{BA_V}
\end{eqnarray}
Here, $^{N}\!E_0\,=\,^{N}\!E_0(n^N_1,n^N_2)$ is the BMF(2) energy of the reference system,  
$n^N_1$ and $n^N_2$ are the corresponding optimal occupation numbers,
$^{N+1}\!E(n^N_1+1,n^N_2)$ and $^{N+1}\!E(n^N_1,n^N_2+1)$ are the MF(2) energies of the states 
where a boson has been attached to the first and to the second fragment, respectively.
In Fig.\ref{fig4}A and Fig.\ref{fig4}B we schematically plot the vertical boson affinities
$BA_V(2)$ and $BA_V(1)$ as the vertical lines connecting the corresponding points on the energy curves.

In the frozen orbital approximation the energies of the states to which the boson 
has been suddenly attached to the $\phi_1$ or $\phi_2$ orbital take on the form 

\alpheqn\begin{eqnarray}
^{N+1}\!E(n_1+1,n_2)\, =\,
(n_1+1) h_{11} + \lambda_0 \frac{(n_1+1)(n_1-0)}{2}\int|\phi_1|^4 d\vec{r}+ & &  \nonumber\\
n_2 h_{22} + \lambda_0 \frac{n_2(n_2-1)}{2}\int|\phi_2|^4 d\vec{r} +
+2 \lambda_0 (n_1+1) n_2 \int|\phi_1|^2 |\phi_2|^2 d\vec{r}  \label{E_N+1.A}  & & \\  & & \nonumber\\
^{N+1}\!E(n_1,n_2+1)\,=\,
(n_2+1) h_{22} + \lambda_0 \frac{(n_2+1)(n_2-0)}{2}\int|\phi_2|^4 d\vec{r}+& & \nonumber\\ 
n_1 h_{11} + \lambda_0 \frac{n_1(n_1-1)}{2}\int|\phi_1|^4 d\vec{r} +   
+2 \lambda_0 n_1 (n_2+1) \int|\phi_1|^2 |\phi_2|^2 d\vec{r}  \label{E_N+1.B}
\end{eqnarray}
\reseteqn
By substructing these energies from the BMF energy of the reference N-particle state $^{N}\!E_0$,
we obtain the vertical boson affinities $BA_V(1)$ and $BA_V(2)$ in the
frozen orbital approximation:
\begin{eqnarray}
BA_V(1)&=&-\mu_{11}-\lambda_0 \int|\phi_1|^4 d\vec{r} \nonumber\\
BA_V(2)&=&-\mu_{22}-\lambda_0 \int|\phi_2|^4 d\vec{r}. 
\label{BA_V_FRZ}
\end{eqnarray}
Interestingly, the vertical boson affinities in the frozen orbital approximation are not fully determined
by the chemical potentials in contrast to our finding for the vertical boson ionization potentials
discussed in the preceding subsection.

It is informative to notice that the energy gained by adding a boson to the non-fragmented condensate
without changing the corresponding orbital $\varphi$ for a finite number of bosons reads: 
\begin{equation}
\,^N\!E_{GP}-\,^{N+1}\!E_{GP}\,=\,-\mu_{GP}-\lambda_0 \int|\varphi|^4 d\vec{r}.
\label{BA_GP}
\end{equation}
Hence, even at the level of the standard GP theory, 
the energy needed to remove a boson from a non-fragmented condensate 
differs from that gained by adding a boson to this condensate.

The above findings reveal differences between bosonic and fermionic systems. 
In contrast to the electron affinity which in the framework 
of the Hartree-Fock mean-field approach is given by the negative of 
the virtual orbital's energy, the boson affinity does not depend on the virtual orbital
and, in addition, is subject to a correction term proportional to $\lambda_0$.

\subsection{Numerical examples}


For illustration purposes we evaluate all the above introduced adiabatic and vertical boson ionization 
potentials and boson affinities for several reference systems made of $N=5,10,25,100$ and 1000 bosons, 
keeping $\lambda=0.9$ fixed throughout. In the GP theory all these systems 
are characterized by the same chemical potential.
Tab.\ref{Tab1} summarizes the computational results on the vertical boson ionization potentials
and vertical boson affinities obtained within the framework of the direct scheme in Eqs.\ref{BIP_V}
and \ref{BA_V}, respectively. 
For comparison we also evaluate all these quantities using the frozen orbital 
approximation as given in Eqs.\ref{BIP_V_FRZ} and \ref{BA_V_FRZ}. In Tab.\ref{Tab1} these approximate numbers
are given in parenthesis. 
By comparing all these quantities we conclude that the frozen orbital 
approximation provides very accurate results for the systems with 
a large number of bosons and surprisingly accurate results for the system with a small number of bosons.

The main physical conclusion derived is that the two-fold fragmented states are characterized by two
different vertical ionization potentials and by two different vertical boson affinities.
In contrast, the non-fragmented GP state is characterized by a single ionization potential
and single affinity. At least in principle, this difference might be used to distinguish 
fragmented and non-fragmented states.
The values of the various vertical boson ionization potentials and vertical boson affinities 
differ from each other clearly for the systems made of a small number of bosons.
As the number of bosons increases, all these values become more similar to each other and also
to the negative of the GP chemical potential.

Closing this section, we would like to mention that 
we used the numerical MF(2) data obtained for the reference system with $N=25$ 
and for the $N\pm1$ systems with 24 and 26 bosons to plot the diagrams presented in Figs.\ref{fig4}.
For all these systems we fixed the interaction strength at $\lambda_0=0.9/(25-1)$.
The optimal occupations of the first and second orbitals at the minimum of the MF(2) energy curve 
of the reference system with 25 bosons are found to be $n^N_1=23.08$ and $n^N_2=1.92$, respectively.
The non-integer occupation numbers appear due to the underlying mean-field approximation. 
They are to be considered as average values of the respective particle-number operators.
We use these occupation numbers as the reference scale for the x-axes in Fig.\ref{fig4}A and Fig.\ref{fig4}B.

\section{Boson transfer energies and the origin of the minima of MF energies}

We define the {\it boson transfer energy} (BTE) as the energy needed to transfer
a boson from one orbital to the other. 
In the framework of the MF(2) theory 
we start with the system which has the energy $^NE(n_1,n_2)$ and evaluate the BTE
as the energy difference between this energy and the energy of the "final" state where 
the occupation number of the first orbital is reduced by one to $n_1-1$ and the occupation number 
of the second orbital is increased by one to $n_2+1$.
In principle, the energy needed to move a boson from the first orbital to the second one 
differs from the energy of the inverse process where
a boson is transferred from the second orbital to the first one.
The $BTE_{(1\rightarrow2)}$ and $BTE_{(1\leftarrow2)}$ are given by:
\begin{equation}
BTE_{(1\rightleftharpoons2)}\,=\,^{N}\!E(n_1\mp1,n_2\pm1)-\,^{N}\!E(n_1,n_2).
\label{BTE}
\end{equation}
In the frozen orbital approximation the boson transfer energies 
can be straightforwardly evaluated using these definitions and the MF(2) expressions
for the energies. The results take on the following form:
\begin{eqnarray}
BTE_{(1\rightarrow2)} & = & \mu_{22}-\mu_{11} 
+\frac{\lambda_0}{2}\int|\phi_2|^4 d\vec{r}-2\,\lambda_0\int|\phi_1|^2|\phi_2|^2 d\vec{r}  \nonumber\\
BTE_{(1\leftarrow2)} & = & 
\mu_{11}-\mu_{22}
+\frac{\lambda_0}{2}\int|\phi_1|^4 d\vec{r}-2\,\lambda_0\int|\phi_1|^2|\phi_2|^2 d\vec{r}  \nonumber\\
\label{BTE_FRZ}
\end{eqnarray}
These results will be used in the discussion below.

It is quite natural to suppose that at equilibrium an exchange of the particles 
between different orbitals (fragments) must be suppressed in the fragmented state.
In other words, the energy needed to transfer a boson from the first orbital to the second one,
is equal to the energy needed for the inverse process - a transition of the boson from the second orbital 
to the first one. 
Therefore, the difference between both boson transfer energies 
\begin{equation}
Q=(BTE_{(1\leftarrow2)}-BTE_{(1\rightarrow2)})/2
\label{Q}
\end{equation}
may serve as a criterion of how far we are from the equilibrium.
Substituting the respective transition energies from Eqs.\ref{BTE_FRZ} 
provides a senseful approximation for $Q$:
\begin{equation}
Q=\mu_{11}-\mu_{22}+\frac{\lambda_0}{2}\int|\phi_1|^4 d\vec{r}-\frac{\lambda_0}{2}\int|\phi_2|^4 d\vec{r}
\label{Q_FRZ}
\end{equation}
To make contact with the MF(2) energy curves and to get deeper insight into 
the origin of the minima of these curves which determine the BMF(2) energy,
it is worthwhile to consider the energy derivative $\frac{d E}{d n_1}$ with 
respect to the occupation number $n_1$.
To evaluate the derivative we use again the frozen orbital approximation 
which has been found to provide accurate results for the boson ionization 
potentials and affinities.
In this approximation we can assume that
the orbitals $\phi_i$ do not depend explicitly on the $n_i$ and 
the direct differentiation of the energy in Eq.\ref{BMF2_energy} with respect to $n_1$ gives:
\begin{equation}
\frac{d E}{d n_1}=\mu_{11}-\mu_{22}+\frac{\lambda_0}{2}\int|\phi_1|^4 d\vec{r}-
\frac{\lambda_0}{2}\int|\phi_2|^4 d\vec{r}
\label{dEdn_FRZ}
\end{equation}
By comparing this derivative with the previous equation for $Q$ one finds that they are identical.

At the optimal orbitals and occupation numbers the MF(2) energy takes on its minimum and 
$\frac{d E}{d n_1}=0$ (see Figs.\ref{fig2A} and \ref{fig2B}).
Consequently, our physical assumption that at the equilibrium the boson transfer energies are equal is 
fully supported by identifying $Q$ with $\frac{d E}{d n_1}$.
Moreover, if the bosonic system is not at equilibrium, then
due to the difference between $BTE_{(1\leftarrow2)}$ and $BTE_{(1\rightarrow2)}$ 
a flow of bosons between the $\phi_1$-manifold and the $\phi_2$-manifold
is enforced until equilibrium is reached. Therefore, the slope of the MF(2) energy curve given
in Eq.\ref{dEdn_FRZ} can be viewed as the "driving force" for the flow of bosons between the two boson subsystems.

To verify these physically appealing results we investigate in more detail 
condensates with $N=25$ and $N=1000$ bosons at $\lambda=0.9$.
The lower panels of Fig.\ref{fig5} show values of $Q$ for different values of $n_1/N$
evaluated via Eq.\ref{Q_FRZ} using the frozen orbital approximation.
Also shown is $\frac{d E}{d n_1}$ obtained by 
numerically differentiating the MF(2) energy curve in Fig.\ref{fig2B}
with respect to $n_1/N$ for the same systems.
On both pictures the values of the optimal occupation numbers $n_1/N$ where the MF(2) 
energy curves take on their minima are marked by vertical dashed lines.
From these figures it is clearly seen that indeed,
at the minimum of the MF(2) energy, the flow of the particles between the fragments is 
completely suppressed, i.e., at the best mean-field state the driving force is equal to zero.

\section{Large N limit} \label{LN}

In the large N limit, $N$ as well as $n_1$ and $n_2$ are much larger than 1
and we may replace $N-1$ by $N$. Repeating the calculations of the preceding sections one finds that all
terms proportional to $\lambda_0 \times 1$ and not to $\lambda_0 \times N$ or 
$\lambda_0 \times n_i$, $i=1,2$, vanish in the large N limit.
Accordingly, we find that in the large N limit not only the vertical boson
ionization potentials are given by the chemical potentials as found in Sec.\ref{BIP_BA} (see Eqs.\ref{BIP_V_FRZ}),
but that also the vertical boson affinities are:
\begin{equation}
BA_V(k)=-\mu_{kk},\,\,\, k=1,2. 
\label{BA_LN}
\end{equation}
Obviously, also the boson affinity in the framework of the GP theory is determined by the 
corresponding GP chemical potential $\mu_{GP}$ in the large N limit.

In the large N limit the boson transfer energy $BTE_{(1\rightarrow2)}$ becomes 
identical to $-BTE_{(1\leftarrow2)}$. The energy needed to transfer a boson in a condensate
with $N$ particles from one fragment to the other one is just given by the difference of the corresponding
potentials:
\begin{equation}
BTE_{(i\rightarrow j)}= \mu_{jj}-\mu_{ii} 
\label{BTE_LN}
\end{equation}
This quantity is then nothing but $BIP_V(i)-BA_V(j)$ which is very appealing for large systems.

In the large N limit the quantity $Q$ introduced in Eq.\ref{Q} and the energy derivative with respect to 
the occupation number $n_1$ become identical and are just determined by the difference of the chemical 
potentials of the fragments. In particular, we find
\begin{equation}
\frac{d E}{d n_1}=\mu_{11}-\mu_{22} 
\label{dEdn_LN}
\end{equation}
which has been derived before in Ref.\cite{BMF2}.

\begin{figure}
\includegraphics[width=11.2cm, angle=-90]{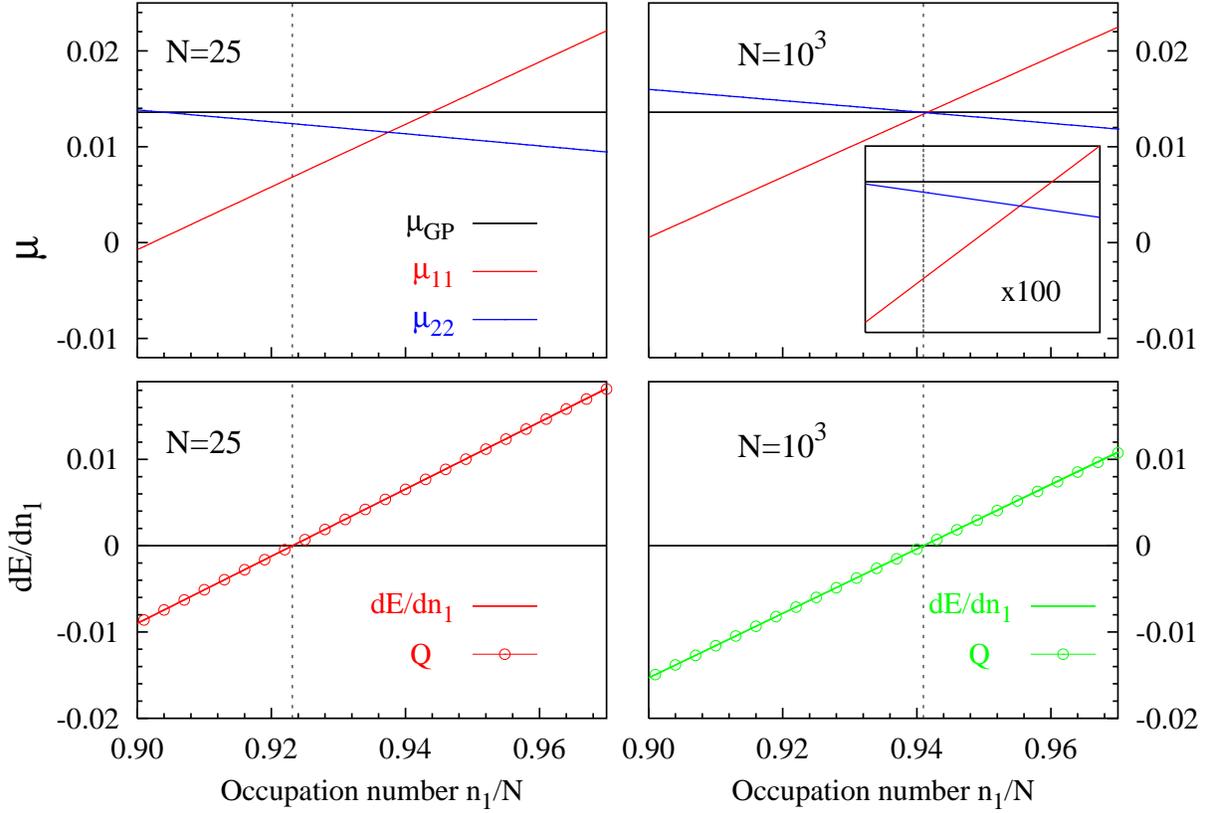}
\caption{(Color online)
The upper two panels show the chemical potentials (diagonal Lagrange multipliers)
$\mu_{11}$ and $\mu_{22}$ for condensates with $N=25$ and $N=1000$ bosons  
as a function of the relative occupation number $n_1/N$.
For comparison, the horizontal solid line shows the value of the corresponding chemical potential
$\mu_{GP}$ obtained in the GP theory.
The inset shows the same curves on a much ($\times10^2$) enlarged scale.
The lower two panels depict the derivative $dE/dn_1$ (solid line) of the total energy
with respect to $n_1$ and the difference between the boson transfer energies $Q$ (points)
for the same systems as in the upper panels. $dE/dn_1$ is obtained by numerical
differentiation of the MF(2) energy curves and $Q$ is evaluated using Eq.\ref{Q_FRZ}.
The vertical dashed lines mark the values of the optimal occupation numbers $n_1/N$
obtained at the minima of the corresponding energy curves. $\lambda=0.9$ is used throughout.
The energy derivatives and all chemical potentials are given in units of $\omega$.
}
\label{fig5}
\end{figure}

To illustrate the impact of the growing number of bosons in condensates
we plot the diagonal Lagrange multipliers $\mu_{11}$ and $\mu_{22}$
as a function of the relative occupation number $n_1/N$
for systems with $N=25$ and $N=1000$ bosons in the upper panels of Fig.\ref{fig5}.
In these panels the values of the optimal occupation numbers $n_1/N$
obtained at the minima of the MF(2) energy curves $E(n_1)$
are marked by vertical dashed lines and the value of the chemical potential
$\mu_{GP}$ obtained in the GP theory by a horizontal solid line.

One can see that for the systems with $N=25$ bosons (see left upper panel of Fig.\ref{fig5})
the crossing point of the $\mu_{11}(n_1/N)$ and $\mu_{22}(n_1/N)$ curves is located 
substantially aside from the vertical dashed line. 
In contrast, the corresponding curves cross the vertical dashed line at almost
the same value of $n_1/N$ for the system with $N=1000$ bosons (see right upper panel of Fig.\ref{fig5}).
However, on an enlarged scale as seen in the inset, the 
exact crossing point of the $\mu_{11}(n_1/N)$ and $\mu_{22}(n_1/N)$ curves  
lies aside from the vertical dashed line marking
the value of the optimal occupation number $n_1/N$.

To better understand these results 
let us analyze Eq.\ref{dEdn_FRZ}. The systems with $N=25$ and $N=1000$ bosons at $\lambda=0.9$ 
are characterized by different values of $\lambda_0=\lambda/(N-1)$.
Therefore, the contributions of the terms which are 
proportional to $\lambda_0$ in Eq.\ref{dEdn_FRZ} are much smaller for the system with $N=1000$
than with $N=25$ bosons where they play an important role. 
In other words, in the large N limit the flow of the bosons 
is determined by the differences of the chemical potentials $\mu_{11}-\mu_{22}$ only. 
Interestingly, $\frac{d E}{d n_1}=0$ holds at the best mean field and consequently the
chemical potentials of the fragments are identical to each other in the ground state
of the condensate in the large N limit.

\section{Conclusions}
In this article condensates with $N$ identical repulsive bosons immersed into a double-well external 
potential have been investigated at different levels of multi-orbital mean-field theories where
optimal orbitals and optimal occupation numbers are determined variationally.

At the level of one-orbital mean-field theory (Gross-Pitaevskii) the ground state of the system
reveals a bifurcation scenario at a critical value of the non-linear parameter 
$\lambda_{cr}=\lambda_0(N-1)=0.837$.
From this $\lambda_{cr}$ on the delocalization of the bosons over the two wells 
of the trap potential becomes energetically more favorable 
than the localization of the BEC in the deeper well. 

The two-orbital mean-field theory predicts the existence of two-fold fragmented states, i.e., 
both orbitals of these states are occupied macroscopically. 
Depending upon the number of particles in the BEC and/or the strength of the interparticle interaction, 
this fragmented state can be the ground state of the system.
For condensates with a small number of particles the fragmented ground state is 
the only stable state, because the non-fragmented state predicted by the GP theory
corresponds to the maximum of the energy in the two-orbital scenario.

By applying the three-orbital mean-field we verify numerically that for 
the double-well external potential studied here the overall best mean field 
is achieved within two orbitals (BMF(2)).
The inclusion of third orbital does not improve the mean-field description
as the variational principle minimizes the energy at a vanishing population of the third orbital.
Hence, the two-fold fragmented state obtained is the physical state of the system.
Once $N\ge N_{max}$, where $N_{max}$ depends on the interparticle interaction strength,
the energy difference between the fragmented and the non-fragmented state may be very small,
indicating that in these cases two physically different states may coexist. 

Inspired by similarities between fragmented states in bosonic systems and 
shell structures in the fermionic world, we introduce vertical and adiabatic boson 
ionization potentials and boson affinities.
In this respect we discuss the chemical potentials of the fragments and their relation 
to boson ionization and attachment measurements.
The energy splitting between the values corresponding to the different fragments 
are small for systems with a large number of bosons while they might be distinguishable
for systems with a small number of bosons.

We also introduced a boson transfer energy as the energy needed to move a boson from 
fragment "1" to fragment "2". We proved that only at equilibrium the 
same energy is needed for the inverse process where a 
boson is transferred from fragment "2" to fragment "1".
It is argued that the difference between both boson transfer energies may serve as 
a criterion of how far we are from equilibrium.
This difference is easily evaluated at each point of the multi-orbital energy surface 
and may speed up the search of the energy minimum.

Explicit expressions have been determined within the frozen orbital approximation
for the boson ionization potentials, boson affinities and boson transfer energies.
These expressions shed light on the physical content of the quantities.
The frozen orbital approximation is based on the 
assumption that adding or removing a single boson from a system with $N$ bosons 
will not strongly change the orbitals of the other bosons and hence these 
orbitals can be assumed unchanged.
We present numerical results which demonstrate that this approximation
is valid for the condensates studied here.

Finally, we would like to stress that the present findings are general
and not at all restricted to the geometry of the double-well trap potential
discussed here. We find similar results for other double- and multi-well trap 
potentials including potentials used in current experiments.

The authors acknowledge useful discussions with Ofir Alon and Kaspar Sakmann.


\pagebreak

\pagebreak
\begin{table}
\caption{The adiabatic and vertical boson ionization potentials $BIP_A$ and $BIP_V(k)$
and boson affinities $BA_A$ and $BA_V(k)$ for condensates with $N=5,10,25,100$ and 1000 bosons.
Shown are quantities evaluated according to their 
definitions in Eqs.\ref{BIP_A},\ref{BIP_V} and \ref{BA_A},\ref{BA_V} as differences of total energies
and compared with those computed using the frozen orbital approximation
in Eqs.\ref{BIP_V_FRZ} and \ref{BA_V_FRZ}. These approximate numbers are given in parenthesis.
$\lambda=0.9$ is used throughout.
}
\begin{tabular}{cccc|ccc}
 \multicolumn{1}{c} { } &
 \multicolumn{3}{c} {\hrulefill\ Ionization \hrulefill\ } &
 \multicolumn{3}{c} {\hrulefill\ Affinity   \hrulefill\ } \\
 \hline
 \multicolumn{1}{c} {N} &
 \multicolumn{1}{c} {$BIP_A$} &
 \multicolumn{1}{c} {$BIP_V(1)$} &
 \multicolumn{1}{c} {$BIP_V(2)$} &
 \multicolumn{1}{c} {$BA_A$} &
 \multicolumn{1}{c} {$BA_V(1)$} &
 \multicolumn{1}{c} {$BA_V(2)$} \\
 \hline
5    & -0.007508 &  0.025694 & -0.006132 & -0.020167& -0.052911 & -0.021267\\
     &           &( 0.027242)&(-0.006248)&          &(-0.055309)&(-0.021819)\\
\hline
10   & -0.010946 &  0.003733 & -0.010390 & -0.016514& -0.031100 & -0.017021\\
     &           &( 0.004460)&(-0.010387)&          &(-0.031993)&(-0.017142)\\
\hline
25   & -0.012613 & -0.007123 & -0.012412 & -0.014692& -0.020167 & -0.014886\\
     &           &(-0.006844)&(-0.012405)&          &(-0.020470)&(-0.014910)\\
\hline
100  & -0.013361 &-0.012032  & -0.013314 & -0.013865& -0.015193 & -0.013912\\
     &           &(-0.011964)&(-0.013311)&          &(-0.015262)&(-0.013916)\\
\hline
1000 & -0.013576 & -0.013445 & -0.013572 & -0.013626& -0.013758 & -0.013631\\
     &           &(-0.013438)&(-0.013571)&          &(-0.013764)&(-0.013631)\\
\hline
& $\mu_{GP}$ &=0.013599 \\
\end{tabular}
\label{Tab1}
\end{table}

\end{document}